\documentclass[12pt]{article}
\usepackage{amsmath}
\usepackage{amssymb}
\begin{document}
\vskip 2cm
\begin{center}
{\sf {\Large  Nilpotent and Absolutely Anticommuting Symmetries in the Freedman-Townsend Model:
Augmented Superfield Formalism}}

\vskip 3.0cm

{\sf A. Shukla $^{(a)}$, S. Krishna$^{(a)}$, R. P. Malik$^{(a,b)}$}\\
$^{(a)}$ {\it Physics Department, Centre of Advanced Studies,}\\
{\it Banaras Hindu University, Varanasi - 221 005, (U.P.), India}\\
$^{(b)}$ {\it DST Centre for Interdisciplinary Mathematical Sciences,}\\
{\it Faculty of Science, Banaras Hindu University, Varanasi - 221 005, India}\\
{\small {\sf {e-mails: ashukla038@gmail.com; skrishna.bhu@gmail.com;   rpmalik1995@gmail.com}}}

\end{center}

\vskip 2cm

\noindent
{\bf Abstract:} 
We derive the off-shell nilpotent and absolutely anticommuting Becchi-Rouet-Stora-Tyutin (BRST) and anti-BRST 
symmetry transformations, corresponding to the (1-form) Yang-Mills (YM) and (2-form) tensorial  
gauge symmetries of the four (3 + 1)-dimensional (4D)
Freedman-Townsend (FT) model, by exploiting the augmented version of Bonora-Tonin's (BT) superfield 
approach to BRST formalism where the 4D flat Minkowskian theory is generalized onto the 
(4, 2)-dimensional supermanifold. 
One of the {\it novel} observations is the fact that we are theoretically compelled to go beyond the 
horizontality condition (HC) to invoke an additional set of gauge-invariant restrictions (GIRs) 
for the derivation of the full set of proper (anti-)BRST symmetries. To obtain the (anti-)BRST symmetry transformations,
corresponding to the tensorial (2-form) gauge symmetries within the framework of augmented version of BT-superfield
approach, we are logically  forced to {\it modify} the FT-model to incorporate an auxiliary 1-form field and the kinetic
term for the antisymmetric (2-form) gauge field. This is also a {\it new} observation in our present investigation.
We point out some of the key differences between the modified FT-model and Lahiri-model 
(LM) of the dynamical non-Abelian 2-form gauge theories. We also briefly mention a few similarities.



\newpage

\section {Introduction}

The (super)string theories (which are at the forefront of research in 
theoretical high energy physics) encompass in their ever-widening folds the quantum 
aspects of gravity as well as gauge theories and, hence, represent one of the promising and 
leading candidates for the unification of {\it all} fundamental interactions of nature (see, e.g. [1-3]). 
The $p$-form ($p = 1, 2, 3,...$) gauge fields are an integral part of the excitation spectrum of 
quantized  (super)string theories. These gauge fields are interesting  
in their own right as far as their associated field theories are
concerned. We have applied the superfield approach to BRST formalism (see, e.g. [4-9])
to study $p = 1, 2, 3$ form (non-)Abelian theories and obtained
their proper (anti-)BRST, (anti-)co-BRST and bosonic symmetries, to prove {\it some} of them to be
the field theoretic models for the Hodge theory [10-15]. The latter field theories, for instance, are: 2D free (non-)Abelian 1-form gauge theories,
2D Abelian 1-form gauge theory in interaction with Dirac fields, 4D free 2-form and 6D free 3-form  Abelian gauge theories.
In fact, the higher $p$-form ($p = 2, 3, ...$) gauge theories have provided a fertile ground 
for the germination of {\it new} theoretical ideas as far as the study of their quantum
field theoretic  aspects is concerned.

There are a couple of widely well-known models for the non-Abelian 2-form gauge theories in the physical
four (3 + 1)-dimensions of spacetime which are topologically massive because of their coupling with
the non-Abelian 1-form gauge theories. These are celebrated Freedman-Townsend (FT) model [16] 
and  Lahiri-model (LM) [17].
In the former case, the non-Abelian 2-form $B^{(2)} = [(dx^\mu \wedge dx^\nu)/2!\, B_{\mu\nu}]$ field $B_{\mu\nu}$
is an {\it auxiliary} field but it is a {\it dynamical} field in the case of the latter 
(because there is a kinetic term
for the 2-form field). The purpose of our present investigation
is to apply the  augmented version of superfield formalism 
(see, e.g. [7-9]) to study the FT-model of massive topological gauge field theory 
and point out some {\it novel} features associated with it. In this context, there are two {\it novel} observations
that are worth pointing out. First, we are theoretically forced to go beyond the horizontality condition 
to invoke some appropriate GIRs to obtain the full set of proper (anti-) BRST symmetries.
 Second, we are logically  compelled to modify the FT-model to incorporate an auxiliary vector field
 and a kinetic term for the 2-form gauge field (see, subsection 4.1) for the application of the above augmented 
superfield formalism [7-9].

In our present investigation, first of all, we exploit the key ingredients of the augmented superfield
formalism to derive the (anti-)BRST symmetries corresponding to the (1-form) 
YM gauge symmetries  by using theoretical tricks
that are distinctly different from the {\it ones} used in our previous work [18]. 
In particular, in our present endeavor, the GIRs on the superfields
 are totally different from our earlier paper [18]. It was a {\it challenging} problem to derive
the proper (anti-)BRST symmetries, corresponding to the non-Yang-Mills (NYM) tensorial gauge symmetries, for the FT-model
within the framework of augmented BT-superfield formalism.
To achieve that goal, in our present investigation, we demonstrate that we are theoretically compelled to write the topological mass term 
of the {\it original} FT-model in a different manner so that we could get a spacetime derivative on the $B_{\mu\nu}$
field. We have accomplished this goal in our present endeavor (see, subsection 4.1) which is 
an important requirement for the application of superfield formalism to derive the NYM symmetries. This artifact enforces
us to incorporate a kinetic term for $B_{\mu\nu}$ field (which is an {\it auxiliary } field in the {\it original}
FT-model) thereby turning it to be a dynamical field.

For any arbitrary $p$-form (non-)Abelian gauge theory, it is a common folklore to incorporate a gauge invariant kinetic
term for the basic $p$-form gauge field. Thus, for the modified version of FT-model, we have to obtain a 
gauge-invariant  curvature tensor $H_{\mu\nu\eta}$ which is derived from the 3-form 
$H^{(3)} = [(dx^\mu \wedge dx^\nu \wedge dx^\eta)/3!]\; H_{\mu\nu\eta}$. A
simple covariant derivative on $B_{\mu\nu}$ does {\it not} do the job.
Thus, we are forced to incorporate an auxiliary field ($K_\mu$) and demand a specific type of (N)YM transformations on it so that
we could obtain a gauge-invariant $H_{\mu\nu\eta}$. This goal has {\it also} been achieved in our present endeavor.
Ultimately, we have obtained a modified version of  Lagrangian density for
the original FT-model which respects YM as well as tensorial (NYM) gauge symmetries. 
The mathematical form of the curvature tensor $H_{\mu\nu\eta}$ is similar to LM with a different definition for
the covariant derivative and curvature tensor for the 1-form gauge field (cf. subsection 4.1).

Even though the appearance of our modified version of Lagrangian density for the FT-model is similar to
the {\it one} for the dynamical non-Abelian 2-form gauge theory of LM,  there are distinct differences in the topological
mass term that are incorporated in both these theories. Furthermore, the covariant derivatives and 
gauge-field curvature tensors are different in both these theories. In the former, the covariant derivative
is defined in terms of both 1-form fields $A_\mu$ and $\phi_\mu$ whereas, in the latter case,
it is w.r.t. $A_\mu$ field {\it only}. One of the important  features
of the modified version of FT-model is the fact that the modified form of the topological mass term remains invariant
under both (1-form) YM and tensorial (NYM) gauge symmetries even though  it looks completely 
different from the original FT-model and LM. This is a completely {\it new} observation in our present investigation
which is very gratifying  as far as symmetry properties of our present theory are concerned.

Our present endeavor is motivated by the following key factors. First, it is urgent for us to apply
the superfield formalism to the description of FT-model because we have performed a similar kind of
analysis for the dynamical non-Abelian 2-form theory (LM) in our previous work [19].
This is essential for the sake of comparison and deep understanding.  Second, our
attempt yields some novel observations in the context of FT-model which enriches our overall insights and 
understanding of the non-Abelian 2-form gauge theory. 
Third, our exercise leads to the derivation of the topological mass term which
looks completely different from LM and it turns out to respect both YM and tensorial (NYM) gauge symmetries. 
Four, the 4D non-Abelian 2-form field has also relevance in the context of 
(super)string theory and related extended objects.
Hence, the study of its field theoretic aspects is important. 
Finally, our present attempt is our modest step towards our main goal of providing a unitary, consistent 
and renormalizable non-Abelian 2-form theory whose precise 
construction, even now, is an outstanding problem in the realm of quantum field theory.

The material of our present investigation is organized as follows. Our Sec. 2 is devoted to a brief 
sketch of the (1-form) YM and (2-form) tensorial gauge symmetries of the FT-model in the Lagrangian formulation. 
In Sec. 3,  we derive the proper (anti-)BRST symmetry transformations (corresponding to the YM symmetries) 
by exploiting our superfield formulation which (as far as the basic inputs are concerned) is completely different from the theoretical
tricks of an earlier work [18]. Our Sec. 4 deals with
the derivation of proper (anti-)BRST symmetries, corresponding to the (2-form) tensorial gauge symmetry,
within the framework of our augmented superfield formalism where we are theoretically compelled to modify the FT-model by incorporating
an auxiliary 1-form vector field  and the kinetic term for the 2-form gauge field (where the curvature tensor for the
latter depends on the former field). Finally, in Sec. 5, we make some concluding remarks.

Our Appendix A deals with some intermediate steps that are needed in the proof of the 
tensorial gauge symmetry invariance of the modified FT-model [cf. (25), (28)].
In our Appendix B, we provide some explicit explanation for the ghost fields, needed in the theory, 
corresponding to the 1-form fields $A_\mu$ and $\phi_\mu$.

{\bf{\it General convention and notations:}} Throughout the whole body of our text, we shall denote 
the (anti-)BRST symmetry transformations by $s_{(a)b}$ corresponding to YM and NYM gauge symmetries.
We shall focus only on the {\it internal} symmetries of the theory and treat the 4D Minkowskian flat
 (i.e. $\eta_{\mu\nu} =$  diag $(+1, -1, -1, -1)$) spacetime in the background with the flat
metric $\eta_{\mu\nu}$. Thus, we shall not discuss about any {\it spacetime} symmetries
of our present 4D theory.

\section{Preliminaries: (Super)gauge and (anti-)BRST \\Symmetries in the Lagrangian Formalism}

Let us begin with the four (3+1)-dimensional (4D) Lagrangian density of the FT-model of topologically massive
gauge theory where the 1-forms ($A^{(1)} = dx^\mu A_\mu \cdot T$, $\phi^{(1)} = dx^\mu \phi_\mu \cdot T$) 
and 2-form [$B^{(2)}= \frac{1}{2!} (dx^\mu \wedge dx^\nu)\, B_{\mu\nu} \cdot T$] gauge fields are merged 
together through the celebrated $(B \wedge {\cal F})$ term. 
The explicit form of the Lagrangian density is\footnote{We adopt here the convention and notations such that
the background 4D flat Minkowskian metric has the signature (+1, -1, -1, -1) and totally
antisymmetric 4D Levi-Civita tensor $\varepsilon_{\mu\nu\eta\kappa}$ is chosen to satisfy 
$\varepsilon_{\mu\nu\eta\kappa} \varepsilon^{\mu\nu\eta\kappa} = - 4!,
\varepsilon_{\mu\nu\eta\kappa} \varepsilon^{\mu\nu\eta\rho} = - 3 ! \delta^\rho_\kappa,$ etc., and $\varepsilon_{0123} = + 1$. 
We also choose the dot and cross products $ P \cdot Q = P^a Q^a, (P \times Q) = f^{abc} P^a Q^b T^c$ in the SU(N)
Lie algebraic space where the generators $T^a$ satisfy $[T^a, T^b] = i f^{abc} T^c$ with
$a, b, c....= 1, 2, ....N^2 - 1$ . Here the structure constants $f^{abc}$ are chosen to be totally antisymmetric
in $a, b, c$ for the semi-simple SU(N) Lie algebra (see, e.g. [20] for details).} (see, e.g. [18])
\begin{eqnarray}
{\cal L}_{(0)} = - \frac{1}{4}\; F^{\mu\nu} \cdot F_{\mu\nu} + \frac{m^{2}}{2} \;\phi^{\mu} \cdot \phi_\mu
+ \frac{m}{4} \;\varepsilon^{\mu\nu\eta\kappa}  B_{\mu\nu} \cdot {\cal F}_{\eta\kappa} ,
\end{eqnarray}
where the 2-form $F^{(2)} = d A^{(1)} + i \;(A^{(1)} \wedge A^{(1)})$ defines the curvature tensor 
$F_{\mu\nu} = \partial_\mu A_\nu - \partial_\nu A_\mu - (A_\mu \times A_\nu)$
for the 1-form gauge field $A_\mu = A_\mu \cdot T$. 
The other curvature tensor is
\begin{eqnarray}
{\cal F}_{\mu\nu} &=& F_{\mu\nu} + f_{\mu\nu} - (A_\mu \times \phi_\nu)
- (\phi_\mu \times A_\nu) \nonumber\\ 
 &\equiv& \partial_\mu (A_\nu + \phi_\nu) -\partial_\nu (A_\mu + \phi_\mu) - (A_\mu + \phi_\mu) \times (A_\nu + \phi_\nu),
\end{eqnarray}
where $f_{\mu\nu} = \partial_\mu \phi_\nu - \partial_\nu \phi_\mu - (\phi_\mu \times \phi_\nu)$
is the curvature tensor for the additional 1-form [$\phi^{(1)} = dx^\mu (\phi_\mu \cdot T) $] 
field $\phi_\mu$ and the 2-form field $B_{\mu\nu}$ is an {\it auxiliary field} in the theory as it has no 
explicit kinetic term. It is clear from (1) that, in the natural units 
($ \hbar = c =1$), the mass dimension of ($A_\mu, \phi_\mu, B_{\mu\nu}$) is {\it one}. As a consequence,
 the parameter $m$, in the Lagrangian density (1), has the dimension of {\it mass} (in our  present 4D gauge theory).

The above Lagrangian density (1) respects ($\delta_g {\cal L}_{(0)} = 0$)
the following local, continuous and infinitesimal (1-form) Yang-Mills (YM) gauge symmetry  
transformations ($\delta_g$)
\begin{eqnarray}
&&\delta_g \phi_\mu = - (\phi_\mu \times \Omega), \qquad \delta_g A_\mu = D_\mu \Omega ,\qquad
\delta_g F_{\mu\nu} = - (F_{\mu\nu} \times \Omega), \nonumber\\ &&
\delta_g {\cal F}_{\mu\nu} = - ({\cal F}_{\mu\nu} \times 
\Omega ), \quad \qquad \delta_g B_{\mu\nu} = - (B_{\mu\nu} \times \Omega),\nonumber\\ &&
\delta_g f_{\mu\nu} = - (f_{\mu\nu}\times \Omega) 
+ (\phi_\mu \times \partial_\nu \Omega) - (\phi_\nu \times \partial_\mu \Omega), 
\end{eqnarray}
where $\Omega = \Omega \cdot T$ is the infinitesimal
SU(N)-valued local YM Lorentz scalar gauge parameter and  the covariant derivative
$D_\mu \Omega = \partial_\mu \Omega - (A_\mu \times \Omega)$. There is yet another 
continuous symmetry in the theory.  We call this symmetry  as the non-Yang-Mills 
(NYM) or tensorial gauge symmetry ($\delta_t$). 
Under this symmetry, the relevant fields of the theory transform as:
\begin{eqnarray}
\delta_t F_{\mu\nu} = \delta_t f_{\mu\nu} = \delta_t \phi_\mu
= \delta_t {\cal F}_{\mu\nu} = 0, \qquad\quad
\delta_t B_{\mu\nu} = -(\tilde D_\mu \Lambda_\nu - \tilde D_\nu \Lambda_\mu ),
\end{eqnarray}
where $ \tilde D_\mu \Lambda_\nu = \partial_\mu \Lambda_\nu 
- (A_\mu \times \Lambda_\nu ) - (\phi_\mu \times \Lambda_\nu )
 \equiv \partial_\mu \Lambda_\nu - (A_\mu + \phi_\mu) \times \Lambda_\nu$
and $\Lambda_\mu = \Lambda_\mu \cdot T$ is an infinitesimal, local and continuous 
Lorentz vector gauge parameter.

It is straightforward to note that the Lagrangian density (1) transforms to a total 
spacetime derivative  (i.e. $\delta_t {\cal L}_0 =  \partial_\mu [(m/2)\,
\varepsilon^{\mu\nu\eta\kappa} \Lambda_\nu \cdot {\cal F}_{\eta\kappa}] $) under the above
(2-form) tensorial gauge  transformations  (4) where the Bianchi identity 
($\tilde D_\mu {\cal F}_{\nu\eta}+\tilde D_\nu {\cal F}_{\eta\mu}+\tilde D_\eta {\cal F}_{\mu\nu}= 0$)
plays an important role in the proof of the symmetry property of the above Lagrangian density (${\cal L}_0$).
We shall see that, in our Sec. 4, we have to modify the original Lagrangian density (1) by 
incorporating an auxiliary 1-form ($K^{(1)} = dx^\mu\, K_\mu$) vector field $K_\mu$ 
and kinetic term ($\frac{1}{12} H^{\mu\nu\eta}\cdot H_{\mu\nu\eta}$) 
for the 2-form ($B_{\mu\nu}$) field so as to obtain the (anti-)BRST symmetry 
transformations\footnote{We shall see that, in the proof of the (anti-)BRST 
invariance, we shall {\it not} use the Bianchi identity.} corresponding to the
tensorial gauge symmetry transformations (4) within the framework of augmented 
version of BT-superfield formalism. The latter would be fruitfully applicable if 
and only if a derivative term (of some variety)  exists for the $B_{\mu\nu}$ field in the
starting Lagrangian density (see, e.g. Sec. 4 for details).

\section {(Anti-)BRST Symmetries Corresponding to the YM Gauge Symmetry: Superfield Approach}

For the paper to be self-contained, we derive here the (anti-)BRST symmetry
transformations, corresponding to the (1-form) YM gauge symmetry
in an {\it alternative} way than the theoretical trick adopted in [18]. It can be readily noted that 
$\delta_g (\phi_\mu \cdot \phi^\mu)=0$,
$\delta_g ({\cal F}_{\mu\nu} \cdot {\cal F}^{\mu\nu})=0$. Thus, when we generalize our 
ordinary 4D theory onto the (4, 2)-dimensional
supermanifold, we demand the following restrictions on the (super)fields:
\begin{eqnarray}
&&\tilde \Phi_\mu(x,\theta,\bar\theta) \cdot \tilde \Phi^\mu(x,\theta,\bar\theta) = 
\phi_\mu (x) \cdot \phi^\mu (x), \nonumber\\
&&\tilde {\cal F}_{MN}(x,\theta,\bar\theta ) \cdot \tilde {\cal F}^{MN}(x,\theta,\bar\theta ) 
= {\cal F}_{\mu\nu}(x) \cdot {\cal F}^{\mu\nu}(x) ,
\end{eqnarray}
where the super 2-form $\tilde {\cal F}^{(2)} = \frac{1}{2!}\; (dZ^M \wedge dZ^N)\; \tilde{\cal F}_{M N}(x,\theta,\bar\theta)$
defines the (anti)symmetric super-curvature tensor $\tilde {\cal F}_{MN}(x,\theta,\bar\theta)$. 
Here the superspace variable $Z^{M} = (x^{\mu},\theta,\bar\theta)$ characterizes 
the (4, 2)-dimensional supermanifold. The explicit form of the super 2-form  is 
\begin{eqnarray}
\tilde {\cal F}^{(2)} = \tilde d\, \left(\tilde A^{(1)} + \tilde \Phi^{(1)}\right) + i \;\left(\tilde A^{(1)} 
+ \tilde \Phi^{(1)}\right) \wedge \left(\tilde A^{(1)} + {\tilde \Phi}^{(1)}\right),
\end{eqnarray}
where $\tilde A^{(1)} = d Z^M (A_M \cdot T)$ and $\tilde \Phi^{(1)} = d Z^M (\Phi_M \cdot T)$
are the super 1-form connections which are the generalizations of the ordinary 1-forms 
$A^{(1)}= dx^{\mu} (A_\mu \cdot T)$ and $\phi^{(1)}= dx^{\mu} (\phi_\mu \cdot T)$, respectively,
and $\tilde d$ (with $\tilde d^{2} = 0$) is the supersymmetric generalization of the 
ordinary exterior derivative $d= dx^{\mu}\;\partial_\mu$ (with $d^{2} = 0$)
onto the (4, 2)-dimensional supermanifold. Within the framework of our augmented
superfield formalism [7-9], we require that the gauge invariant quantities 
should be independent of the Grassmannian variables (i.e. ``soul'' coordinates) $\theta$ and $\bar\theta$. This
is why, we have taken the GIRs in  the above equation (5).

The above generalizations, from the ordinary 4D Minkowskian spacetime manifold onto 
the (4, 2)-dimensional supermanifold, can be explicitly expressed as\footnote{ It should be noted that we have
taken the generalization of (anti-)ghost fields $(\bar C)C$ onto the 
(4, 2)-dimensional supermanifold {\it only} in the expansion of
$\tilde A^{(1)}(x,\theta,\bar\theta)$ but {\it not} in $\tilde \Phi^{(1)}(x,\theta,\bar\theta)$. 
This is due to the fact that there 
are {\it only} one set of (anti-)ghost fields in the theory which are associated with the 
1-form ($A^{(1)} = dx^\mu (A_\mu \cdot T)$) YM gauge potential $A_\mu \equiv A_\mu \cdot T$.
We have discussed about this issue in our Appendix {\bf B} for clarification
(where similar ghost-structure has been taken into account for the 1-form fields). }  
(see, e.g. [18]):
\begin{eqnarray}
d \rightarrow  \tilde d &=& dx^{\mu}\;\partial_\mu + d\theta\;\partial_{\theta} + d\bar\theta\;\partial_{\bar\theta},\nonumber\\
A^{(1)} \rightarrow \tilde A^{(1)} &=& dx^{\mu}\;\tilde{\cal B}_\mu (x, \theta, \bar\theta)+ d\theta\;{\tilde {\bar{\cal F}} }(x, \theta, \bar\theta) 
+ d \bar\theta\;\tilde {\cal F} (x, \theta, \bar\theta),\nonumber\\
\Phi^{(1)} \rightarrow \tilde \Phi^{(1)} &=& dx^{\mu} \;\tilde \Phi_\mu (x, \theta, \bar\theta),
\end{eqnarray}
where $ \tilde{\cal B}_\mu (x,\theta,\bar\theta), 
{\tilde {\bar{\cal F}}} (x, \theta, \bar\theta), \tilde{{\cal F}} (x, \theta, \bar\theta), \Phi_\mu (x, \theta, \bar\theta)$
are the superfields  corresponding to the 4D local fields $A_\mu(x), C(x), \bar C(x), \phi_\mu (x)$
of the 4D (anti-)BRST invariant local field theory. This becomes explicit from the following expansions (see, e.g. [18]):
\begin{eqnarray}
\tilde {\cal B}_\mu (x, \theta, \bar\theta) &=& A_\mu (x) + \theta \;{\bar R}_\mu (x) + \bar\theta \; R_\mu (x) 
 + i\; \theta \,\bar\theta\; P_\mu (x),\nonumber\\
\tilde \Phi_\mu (x, \theta, \bar\theta) &=& \phi_\mu (x) + \theta \; {\bar S}_\mu (x) + \bar\theta \;S_\mu (x)  
+i \;\theta\, \bar\theta \;T_\mu (x), \nonumber\\ 
\tilde {\cal F}(x, \theta, \bar\theta) &=& C(x) + \theta \; {\bar B_1} (x) + \bar\theta \;B_1 (x) 
 +i \;\theta \bar\theta \;s (x), \nonumber\\
{\tilde {\bar {\cal F}}}(x, \theta, \bar\theta ) &=& \bar C(x) + \theta \; {\bar B}_2 (x) + \bar\theta \;B_2 (x)  
+i \;\theta\, \bar\theta \;{\bar s}(x).
\end{eqnarray}
In the above, the limiting case $(\theta, \bar\theta) = 0$, leads to retrieving of the 4D local fields from the superfields.
On the r.h.s., we have ($R_\mu, \bar R_\mu, s,\bar s, S_\mu, \bar S_\mu$) as the fermionic and 
($P_\mu, T_\mu, B_1, \bar B_1, B_2, \bar B_2$) as the bosonic secondary fields.
 In addition, we require the generalization of the auxiliary field $B_{\mu\nu}(x)$ onto the 
(4, 2)-dimensional supermanifold as  $\tilde {\cal B}_{\mu\nu}(x,\theta,\bar\theta)$ which has the following expansion along the 
Grassmannian ($\theta,\bar\theta$) directions:
\begin{eqnarray}
\tilde {\cal B}_{\mu\nu}(x, \theta, \bar\theta) &=& B_{\mu\nu} (x) + \theta \;{\bar R}_{\mu\nu} (x) + \bar\theta \; R_{\mu\nu} (x) 
 + i\; \theta\, \bar\theta\; S_{\mu\nu} (x),
\end{eqnarray}
where $R_{\mu\nu}$ and ${\bar R}_{\mu\nu}$ are the fermionic secondary fields  and $S_{\mu\nu}$ is a
 bosonic secondary field. It is self-coincident that the fields $R_{\mu\nu}, {\bar R}_{\mu\nu}$
 and $S_{\mu\nu}$ are {\it all} antisymmetric in $\mu$ and $\nu$
because the basic field $B_{\mu\nu}$ is antisymmetric.

The analogue of HC of the usual (non-)Abelian 1-form gauge theory [4,5], 
requires that the Grassmannian  components of the following super curvature 2-form:
\begin{eqnarray}
\tilde {\cal F}^{(2)} &=& \frac{1}{2!}\; (dZ^M \wedge dZ^N)\; \tilde{\cal F}_{MN}
\nonumber\\ &\equiv&
\tilde d \,(\tilde  A^{(1)} + \tilde {\cal \phi}^{(1)})  +i\; (\tilde A^{(1)} 
+ \tilde {\cal \phi}^{(1)}) \wedge (\tilde A^{(1)} + {\tilde {\cal \phi}}^{(1)}),
\end{eqnarray}
has to be set equal to zero. For this purpose, one has to compute the 
accurate expansion for the r.h.s. of $\tilde {\cal F}^{(2)}$ in (10).  
This expression can be explicitly written as:
\begin{eqnarray}
&& \tilde{\cal F}^{(2)}=  \frac{1}{2!} {(dx^{\mu}\wedge dx^{\nu})}
  \Bigl ( \partial_\mu (\tilde {\cal B}_\nu + \tilde \Phi_\nu ) 
- \partial_\nu (\tilde {\cal B}_\mu + \tilde \Phi_\mu )
+ i [\tilde {\cal B}_\mu + \tilde  \Phi_\mu , \tilde {\cal B}_\nu 
+ \tilde \Phi_\nu ] \Bigr )\nonumber\\
&&+ (dx^{\mu}\wedge d\theta) \Bigl ( \partial_\mu {\tilde {\bar {\cal F}}}
- \partial_\theta ( \tilde {\cal B}_\mu 
+ \tilde  \Phi_\mu) + i \;[ \tilde {\cal B}_\mu 
+ \tilde \Phi_\mu , \tilde {\bar {\cal F}}] \Bigr )
+ (dx^{\mu}\wedge d\bar\theta ) \Bigl (\partial_\mu \tilde {\cal F }\nonumber\\ 
&&- \partial_{\bar\theta} ( \tilde {\cal B}_\mu 
+ \tilde \Phi_\mu)  + i \;[ \tilde {\cal B}_\mu 
+ \tilde \Phi_\mu , \tilde {\cal  F}] \Bigr )
- (d\theta \wedge d\bar\theta ) \Bigl ( \partial_{\theta} \tilde {\cal F} 
+ \partial_{\bar\theta} {\tilde {\bar {\cal F}}}
+ i \;\lbrace \tilde {\cal F}, \tilde {\bar {\cal F}}\rbrace \Bigr ) \nonumber\\
&&- (d\theta \wedge d\theta ) \Bigl ( \partial_{\theta} {\tilde {\bar {\cal F}}}    
+ \frac {i}{2} \; \lbrace {\tilde {\bar{\cal F}}} , {\tilde {\bar {\cal F}}}\rbrace \Bigr ) 
- (d\bar\theta \wedge d\bar\theta )
\Bigl (\partial_{\bar\theta} \tilde {\cal F} + \frac{i}{2}\; \lbrace \tilde {\cal F}, 
\tilde {\cal F}\rbrace \Bigr ) .
\end{eqnarray}
On equating the coefficients of $ (dx^{\mu}\wedge d\theta),(dx^{\mu}\wedge 
d \bar\theta),\, (d\theta \wedge d\theta ),\,(d \bar\theta \wedge d \bar\theta ),\,
 (d\theta \wedge d\bar\theta )$ equal to zero, we obtain the following {\it five} relationships:
\begin{eqnarray}
 \partial_\mu {\tilde {\bar {\cal F }}} - \partial_\theta ( \tilde {\cal B}_\mu 
+ {\tilde \Phi}_\mu)  + i \;[ \tilde {\cal B}_\mu 
+ {\tilde \Phi}_\mu , {\tilde {\bar {\cal F}}}] = 0, \nonumber\\
 \partial_\mu \tilde {\cal F }- \partial_{\bar\theta} ( \tilde {\cal B}_\mu 
+ {\tilde \Phi}_\mu )  + i [ \tilde {\cal B}_\mu 
+ \tilde \Phi_\mu , \tilde {\cal  F}] = 0, \nonumber\\
\partial_{\theta} {\tilde {\bar{\cal F}}} 
+ \frac {i}{2}\; \lbrace {\tilde {\bar{\cal F}}}, {\tilde {\bar{\cal F}}} \rbrace  = 0,
\quad \partial_{\bar\theta} \tilde {\cal F} 
+ \frac {i}{2} \; \lbrace \tilde {\cal F},{\tilde  {\cal F}} \rbrace = 0,\nonumber\\
 \partial _\theta {\tilde {\cal F}} + \partial_{\bar\theta} {\tilde {\bar{\cal F}}}
+ i \; \lbrace \tilde {\cal F}, {\tilde {\bar {\cal F}}} \rbrace =0.
\end{eqnarray}
However, the above relations (existing for the set of superfields (cf. 8)) are not good 
enough for the derivations of {\it all} the secondary fields (of the expansions (8)) in terms of
the basic and auxiliary fields of the ordinary 4D (anti-)BRST invariant Lagrangian density.

The additional gauge invariants restrictions (GIRs) are {\it now} required to supplement 
(and compliment) the analogue  of the HC (12) to obtain all the set of secondary fields 
in terms of the basic and auxiliary fields of the 4D local version of (anti-)BRST invariant theory.
The top relationship (which is a GIR) of equation (5) serves this purpose. 
In fact, taking the help of expansions in (8), it can be checked that 
$\tilde \Phi_\mu (x, \theta, \bar \theta )\cdot \tilde \Phi^\mu (x, \theta, \bar \theta )
=  {\phi}_\mu(x) \cdot {\phi}^\mu (x)$, leads to the following very important relationships
\begin{eqnarray}  
S_\mu \cdot \phi^{\mu}=0, \quad\qquad \bar S_\mu \cdot \phi^{\mu}=0, 
\quad\qquad  i\;\phi_{\mu} \cdot T^{\mu}= - S_\mu \cdot \bar S^{\mu}.
\end{eqnarray}
At this stage, it can be checked that the top two relations of (12) lead to 
(among other important relations) the following 
\begin{eqnarray}
\bar R_\mu + \bar S_\mu &=& D_\mu \bar C - (\phi_\mu \times \bar C),\qquad
R_\mu + S_\mu = D_\mu C - (\phi_\mu \times C),\nonumber\\
P_\mu + T_\mu &=& D_\mu B_2 - (\phi_\mu \times B_2) + i\;(R_\mu + S_\mu)\times \bar C \nonumber\\
&\equiv& - D_\mu \bar B_1 + (\phi_\mu \times \bar B_1) - i\;(\bar R_\mu + \bar S_\mu)\times C.
\end{eqnarray}
The comparison between (13) and (14) leads to 
\begin{eqnarray}
&& R_\mu = D_\mu C, \;\quad \bar R_\mu = D_\mu \bar C,\;\quad S_\mu = - (\phi_\mu \times C), 
\quad \bar S_\mu = - (\phi_\mu \times{\bar C}),\nonumber\\
&& T_\mu = - (\phi_\mu \times B_2) - i\;(\phi_\mu \times C) \times \bar C 
\equiv  (\phi_\mu \times \bar B_1) - i\;(\phi_\mu \times \bar C) \times C.
\end{eqnarray}
The final expressions, for the secondary fields  that emerge from (12) and (13), are
\begin{eqnarray}
&& R_\mu = D_\mu C, \quad \bar R_\mu = D_\mu \bar C,\quad S_\mu = - (\phi_\mu \times C), 
\quad \bar S_\mu = - (\phi_\mu \times{\bar C}),\nonumber\\
&& T_\mu = - (\phi_\mu \times B) - i\;(\phi_\mu \times C) \times \bar C 
\equiv  (\phi_\mu \times \bar B) - i\;(\phi_\mu \times \bar C) \times C, \nonumber\\
&& P_\mu = D_\mu B + i\;(D_\mu C  \times \bar C) \equiv - D_\mu \bar B - i\;(D_\mu \bar C  \times C), \nonumber\\
&&  B = - \frac {i}{2}\;(C \times C),\;\quad s = -(\bar B \times C),\;\quad  \bar s = (B \times \bar C),\nonumber\\
&&  \bar B = - \frac {i}{2}\;(\bar C \times \bar C), \qquad\quad
 B + \bar  B = - i\;(C \times \bar C),
\end{eqnarray}
where we have identified: ${\bar B}_1 = \bar B, \; B_2 = B$.
Thus, all the secondary fields in the expansions of $\tilde {\cal B}_\mu, 
\tilde {\cal F},\tilde {\bar {\cal F}}$ and $\tilde \Phi_\mu$ have been 
determined in terms of the basic and auxiliary fields of the 4D local 
(anti-)BRST invariant theory. One of the key signatures of a non-Abelian 
theory (within the framework of the BRST formalism) is the existence of the Curci-Ferrari
(CF) condition ($B + \bar B + i\, (C \times \bar C) = 0$) which is present in (16).

The above (anti-)BRST invariant (i.e. $s_{(a)b}\,[B + \bar B + i\, 
(C \times \bar C)] = 0$) CF-condition emerges when we set the coefficient 
of ($d\theta \wedge d\bar\theta$)  equal to zero. We would like to lay 
emphasis on the fact that one of the key signatures of a gauge theory, within the framework
of the BRST formalism, is the existence of the (anti-)BRST invariant CF-type condition. 
This observation is as {\it significant}
as the characterization of a gauge theory in terms of the existence of first-class constraints 
in the language of Dirac's prescription. It is trivial to note that the Abelian 1-form gauge theory is 
characterized by the existence of a {\it trivial} CF-type condition ($B + \bar B = 0$) within the framework of
BRST formalism. We have also been able to demonstrate the mathematical origin for the CF-type
condition in the language of gerbes (see, e.g. [21,22] for details) which formally proves the independent existence
of the nilpotent anti-BRST symmetry (and corresponding conserved charge) {\it vis-\`{a}-vis}  
the nilpotent BRST symmetry (and corresponding conserved BRST charge).

As a consequence of the relations (16), it can be readily seen that we have the 
following explicit expansions of the superfields:
\begin{eqnarray}
 \tilde B_\mu ^{(h,g)}(x, \theta, \bar\theta) &=& A_\mu (x) + \theta \,(D_\mu\bar C (x)) 
+ \bar\theta \,(D_\mu C(x)) 
+ \theta \,\bar\theta \,\Bigl[ i\, D_\mu B - (D_\mu C \times \bar C)(x)\Bigr ],\nonumber\\
&\equiv& A_\mu (x) + \theta \,(s_{ab}\, A_\mu (x)) + \bar\theta \,(s_{b} \,A_\mu (x))  
+ \theta\, \bar\theta \,(s_b \,s_{ab}\, A_\mu (x)) ,\nonumber\\
\tilde \Phi_\mu^{(h,g)} (x, \theta, \bar\theta) &=& \phi_\mu (x) 
+ \theta \,\Big[-(\phi_\mu \times \bar C)(x)\Big]
+ \bar\theta\, \Bigl[-(\phi_\mu \times C)(x)\Bigr ] 
+ \theta\, \bar\theta \,\Bigl [-i(\phi_\mu \times B) \nonumber\\   
&+& (\phi_\mu \times C) \times \bar C \Bigr ](x), \nonumber\\ 
&\equiv& \phi_\mu (x) + \theta \,(s_{ab}\,\phi_\mu (x)) + \bar\theta\,(s_{b}\,\phi_\mu (x))  
 + \theta\, \bar\theta\,(s_b \,s_{ab}\,\phi_\mu (x)) ,\nonumber\\
\tilde {\cal F}^{(h,g)}(x, \theta, \bar\theta) &=& C(x) + \theta\, (i\,\bar B(x))  
+ \bar\theta \,\Bigl [\frac{1}{2}\,(C \times C)(x)\Bigr ] 
+ \theta\, \bar\theta\, \Bigl [-i\,(\bar B \times C)(x)\Bigr ], \nonumber\\
 &\equiv& C(x) + \theta\, (s_{ab}\, C(x)) + \bar\theta \,(s_{b}\, C(x))  
 + \theta\, \bar\theta\, (s_b\, s_{ab}\,C(x)) ,\nonumber\\
{\tilde {\bar {\cal F}}}^{(h,g)}(x, \theta, \bar\theta ) &=& \bar C(x) 
+ \theta\, \Bigl [\frac{1}{2}(\bar C \times \bar C)(x) \Bigr ]
+ \bar\theta \,(i \,B (x)) 
+ \theta \,\bar\theta\, \Bigl [i\,(B \times \bar C)(x)\Bigr ],\nonumber\\
&\equiv& \bar C(x) + \theta \,(s_{ab}\,\bar C(x)) + \bar\theta \,(s_{b} \,\bar C(x))  
 + \theta\, \bar\theta \,(s_b\, s_{ab}\, \bar C(x)),
\end{eqnarray}
where the superscripts $(h,g)$ denote the expansion of the superfields after the application
of HC and GIRs (cf. (12), (5)).
The above expansions establish the mappings\footnote{We precisely mean by notation
$s_b\leftrightarrow  \lim_{\theta \rightarrow 0} \,\frac{\partial}{\partial\theta}$ in (18) as: 
$\frac{\partial}{\partial\theta}\, \Omega^{(h,g)} 
(x, \theta,\bar\theta)\mid _{\theta = 0} \leftrightarrow  s_b\, \Omega(x)$ 
where  $\Omega^{(h,g)} (x, \theta,\bar\theta)$ is the generic superfield obtained after 
the application of HC and GIRs and 
$\Omega(x)$ is the 4D generic ordinary field of the Lagrangian density 
of the 4D (anti-)BRST invariant theory.}
\begin{eqnarray}
&& s_b \;\longleftrightarrow \;\lim_{\theta \to 0}  \frac{\partial}{\partial \bar\theta}, \qquad\qquad
s_{ab} \;\longleftrightarrow \;\lim_{\bar \theta \to 0} \; \frac{\partial}{\partial \theta}, \nonumber\\
&& (s_b \,s_{ab} + s_{ab}\, s_b) \,\quad\longleftrightarrow \,\quad \Bigl(\frac{\partial}
{\partial \bar\theta} \,\frac{\partial}{\partial \theta} 
+  \frac{\partial}{\partial \theta}\,\frac{\partial}{\partial \bar\theta} \Bigr ),
\end{eqnarray}
which demonstrate the automatic nilpotency ($s_{(a)b}^{2} = 0$) and absolute anticommutativity 
($s_b\, s_{ab} + s_{ab} \,s_b = 0$) of the 
(anti-)BRST symmetry transformations $s_{(a)b}$ because it is straightforward to note that
 $(\lim_{\bar\theta \to 0} {\partial_\theta}) ^{2} = 0$, $ (\lim_{\theta \to 0}  
{\partial_{\bar\theta}})^{2} = 0$ and 
$ {\partial_\theta}\,{\partial_{\bar\theta}} + {\partial_{\bar\theta}}\,{\partial_\theta} = 0$.

We have still not been able to deduce precisely  the proper (anti-)BRST symmetry 
transformations for the auxiliary field $B_{\mu\nu}(x)$.
This can be derived  from yet another GIR on the (super)fields of the 
(4, 2)-dimensional supermanifold as illustrated below
\begin{eqnarray}
\tilde {\cal B}_{\mu\nu}(x, \theta, \bar\theta)\cdot\tilde {\Phi_\eta}^{(h,g)}
(x, \theta, \bar\theta) =  B_{\mu\nu}(x)\cdot{\phi}_\eta(x).
\end{eqnarray}
The above choice have been made because it can be checked that 
$\delta_g (B_{\mu\nu} \cdot \phi_\eta) = 0$ which shows the gauge 
invariance of the combination $(B_{\mu\nu} \cdot \phi_\eta)(x)$. The substitution  from (17) and (9)
leads to the following relationships:
\begin{eqnarray}
&& \bar R_{\mu\nu} = -(B_{\mu\nu} \times \bar C), \qquad\quad R_{\mu\nu}
 = - (B_{\mu\nu} \times C), \nonumber\\ 
&& S_{\mu\nu} = -i\;(B_{\mu\nu} \times C) \times \bar C - (B_{\mu\nu} \times B),
\end{eqnarray}
which implies the following expansion
\begin{eqnarray}
&&\tilde B_{\mu\nu} ^{(h,g)}(x, \theta, \bar\theta) = B_{\mu\nu} (x) 
+ \theta\; [-(B_{\mu\nu} \times \bar C)(x)]
 + \bar\theta \; [-(B_{\mu\nu} \times  C)(x)] \nonumber\\
 &&+ \theta \bar\theta\; [(B_{\mu\nu}\times C)\times \bar C - i (B_{\mu\nu}\times B) ](x),\nonumber\\
&&\equiv B_{\mu\nu} (x) + \theta (s_{ab}B_{\mu\nu} (x)) + \bar\theta \;(s_{b}B_{\mu\nu} (x)) 
 + \theta \bar\theta\; (s_b s_{ab}B_{\mu\nu} (x)).
\end{eqnarray}
Thus, ultimately, we have the following (anti-)BRST symmetry transformations 
\begin{eqnarray}
&& s_{ab} A_\mu = D_\mu {\bar C}, \quad s_{ab} {\bar C} = \frac {1}{2} (\bar C \times \bar C), 
\quad s_{ab} C = i\bar B, \quad s_{ab} B = -(B \times \bar C),\nonumber\\  
&& s_{ab} {\bar B} = 0, \quad\qquad s_{ab} \phi_\mu = -(\phi_\mu \times \bar C), \quad\qquad
s_{ab} B_{\mu\nu}= -(B_{\mu\nu} \times \bar C),\nonumber\\
&& s_b A_\mu = D_\mu C,  \quad s_b C = \frac {1}{2}(C \times C),\quad s_b \bar C = i B,
 \quad s_b {\bar B} = -(\bar B \times C), \nonumber\\ 
&& s_b B = 0,\;\;\quad\qquad s_b \phi_\mu = -(\phi_\mu \times C),\;\;\quad
 \qquad s_b B_{\mu\nu} = - (B_{\mu\nu} \times C),
\end{eqnarray}
which are nilpotent of order two (i.e. $s_{(a)b}^{2} = 0 $) and absolutely 
anticommuting in nature  (i.e. $ s_b s_{ab} + s_{ab} s_b = 0$).
The latter property is valid only due to the Curci-Ferrari condition 
$B + \bar B + i(C \times \bar C) = 0 $ which is present in our equation (16).
To be precise, our whole 4D (anti-)BRST invariant theory is defined on a hyper-surface in the 4D flat Minkowskian 
spacetime which is described by the CF-field equation $B + \bar B + i(C \times \bar C) = 0 $.

We note that (anti-)BRST transformations of the Nakanishi-Lautrup auxiliary fields 
($B,\bar B$), in the above, have been 
derived from the requirements of the nilpotency and absolute anticommutativity properties. 
We lay emphasis on the fact that our method
of derivation of the (anti-)BRST symmetries corresponding to the (1-form) 
YM gauge symmetries [cf. (3)] is {\it totally}
different from the method adopted in our earlier work (see, e.g. [18] for details).
 Finally, we mention that the following nilpotent and absolutely anticommuting (anti-)BRST
symmetry transformations for the curvature tensors of our theory, namely;
\begin{eqnarray}
&& s_b F_{\mu\nu}= - (F_{\mu\nu} \times C), \quad 
s_b {\cal F}_{\mu\nu}= - ({\cal F}_{\mu\nu} \times C), \nonumber\\
&& s_b f_{\mu\nu}= - (f_{\mu\nu} \times C) + \phi_\mu \times \partial_\nu C 
- \phi_\nu \times \partial_\mu C,\nonumber\\
&& s_{ab} F_{\mu\nu}= - (F_{\mu\nu} \times \bar C),\quad s_{ab} {\cal F}_{\mu\nu}
= - ({\cal F}_{\mu\nu} \times \bar C),\nonumber\\
 && s_{ab} f_{\mu\nu}= - (f_{\mu\nu} \times {\bar C}) 
+ \phi_\mu \times \partial_\nu {\bar C} - \phi_\nu \times \partial_\mu {\bar C}, \nonumber\\
&& s_b s_{ab} F_{\mu\nu}= - (F_{\mu\nu} \times C) \times \bar C 
- i (F_{\mu\nu} \times B),\nonumber\\ && s_b s_{ab} {\cal F}_{\mu\nu}
= - ({\cal F}_{\mu\nu} \times C)\times \bar C - i ({\cal F}_{\mu\nu} \times B), \nonumber\\
&& s_b s_{ab} f_{\mu\nu} = (f_{\mu\nu} \times C)\times \bar C 
- (\phi_\mu \times \partial_\nu C)\times \bar C
+ (\phi_\nu \times \partial_\mu C)\times \bar C \nonumber\\ &&- i\, (f_{\mu\nu} \times B) 
 - (\phi_\mu \times C) \times \partial_\nu \bar C
+ (\phi_\nu \times C) \times \partial_\mu \bar C \nonumber\\ 
&&+ i (\phi_\mu \times \partial_\nu B) - i (\phi_\nu \times \partial_\mu B),
\end{eqnarray}
are also {\it true} due to the (1-form) YM transformations (22) on 
the basic gauge fields $A_\mu$ and $\phi_\mu$.
This can be checked by using the definitions of 
$f_{\mu\nu},\; F_{\mu\nu}$ and ${\cal F}_{\mu\nu}$ (cf. Sec. 2).

We close this section with the following remarks. First, one can obtain
the coupled (but equivalent) (anti-)BRST invariant Lagrangian densities,
corresponding to the {\it proper} (anti-)BRST symmetry transformations
(22), by exploiting the standard tricks of BRST formalism. This has been 
accomplished in our earlier work  for LM (see, e.g. [18] for details). 
Second, the generators of transformations (22) can be obtained by exploiting
the Noether theorem. These (anti-)BRST charges (and their novel features) 
have been obtained and discussed in our earlier work [18]. 
Finally, the ghost-scale symmetry in the theory can be discussed 
in a straightforward  manner which leads to the existence of a ghost charge. 
We have been able to show the existence of the standard 
BRST algebra in our earlier work [18]. In the forthcoming section, 
we shall discuss about the (NYM)  tensorial gauge symmetry
and corresponding proper (i.e. nilpotent and absolutely  anticommuting) (anti-)BRST symmetries.

\noindent
\section{Tensorial (anti-)BRST Symmetry Transformations: Superfield Formalism}

In this section, first of all, we modify the FT-Lagrangian (1) in order to 
derive the off-shell nilpotent (anti-)BRST symmetry transformations
corresponding to the tensorial (2-form) gauge transformations (4) in our subsection 4.1. 
This modification allows us
to derive the {\it proper} (anti-)BRST symmetry transformations by 
exploiting the tricks and techniques of the augmented
version of BT-superfield formulation in our subsection 4.2.

\subsection{Modified Version of FT-Model}

As pointed out earlier, the 2-form field $B_{\mu\nu}(x)$ is 
{\it an auxiliary} field in the original FT-model [cf. (1)]. 
This observation does {\it not} allow us to apply the basic 
techniques of BT-superfield formulation to derive
the (anti-)BRST symmetry transformations (for the Lagrangian density (1)) 
corresponding to the tensorial (2-form)
{\it gauge} symmetry transformations (4) because we can {\it not} apply the celebrated HC. 
Thus, we modify the Lagrangian density (1) in the following fashion:
\begin{eqnarray}
{\cal L}^{(0)}_{(M)} &=& - \frac{1}{4}\, F^{\mu\nu}\cdot F_{\mu\nu} 
+ \frac{m^2}{2} \phi_\mu \cdot \phi^\mu -
\frac{m}{2}\, \varepsilon^{\mu\nu\eta\kappa} (A_\nu + \phi_\nu)
 \cdot \Big[{\tilde D}_\mu B_{\eta\kappa} \nonumber\\
&+& \frac{1}{2} \,(A_\mu + \phi_\mu) \times B_{\eta\kappa} \Big],
\end{eqnarray}
where we have dropped the total spacetime derivative term from the
 topological mass term 
[$(m/4)\, \varepsilon^{\mu\nu\eta\kappa}\, {\cal F}_{\mu\nu}\cdot B_{\eta\kappa}$] 
of the Lagrangian density (1)
of the {\it original} FT-model so as to get a covariant derivative term on the antisymmetric
 tensor field $B_{\mu\nu}$. This observation
theoretically compels us to incorporate the kinetic term for this field in our
 theory, too. Thus, we propose the following modified 
Lagrangian density ${\cal L}_{(M)}$ corresponding  to the {\it original} Lagrangian density (1) 
of the FT-model, namely; 
\begin{eqnarray}
{\cal L}_{(M)} &=& \frac{1}{12}\, H^{\mu\nu\eta}\cdot H_{\mu\nu\eta}
- \frac{1}{4}\; F^{\mu\nu}\cdot F_{\mu\nu} 
+ \frac{m^2}{2} \,\phi_\mu \cdot \phi^\mu \nonumber\\
 &-&\frac{m}{2}\, \varepsilon^{\mu\nu\eta\kappa} (A_\nu 
+ \phi_\nu) \cdot \Big[ {\tilde D}_\mu B_{\eta\kappa}
+ \frac{1}{2} (A_\mu + \phi_\mu) \times B_{\eta\kappa} \Big],
\end{eqnarray}
where the curvature 3-form ($H^{(3)} = [dx^\mu \wedge dx^\nu
 \wedge dx^\eta/3!]\, H_{\mu\nu\eta}$) defines the curvature
tensor for  antisymmetric non-Abelian antisymmetric tensor 
gauge ($B_{\mu\nu}$) field as
\begin{eqnarray}
H_{\mu\nu\eta} &=& {\tilde D}_\mu B_{\nu\eta} + {\tilde D}_\nu B_{\eta\mu} 
+ {\tilde D}_\eta B_{\mu\nu} - (K_\mu \times {\cal F}_{\nu\eta}) \nonumber\\
&-& (K_\nu \times {\cal F}_{\eta\mu})   - (K_\eta \times {\cal F}_{\mu\nu}).
\end{eqnarray}
In the above, the 1-form ($K^{(1)} = dx^\mu\, K_\mu \cdot T$) field 
$K_\mu$ is a compensating auxiliary field so as to make the curvature 
tensor as a gauge-invariant quantity (i.e. $\delta_t\, H_{\mu\nu\eta} = 0$).
It is elementary to check that under the following tensorial (2-form) gauge transformations
\begin{eqnarray}
&&\delta_t\, B_{\mu\nu} = -({\tilde D}_\mu \Lambda_\nu - {\tilde D}_\nu \Lambda_\mu ), \;\;\qquad 
 \delta_t\, K_\mu =  -\Lambda_\mu, \nonumber\\
 &&\delta_t\, A_\mu = \delta_t\, \phi_\mu = \delta_t\, {\cal F}_{\mu\nu} = \delta_t\,  F_{\mu\nu} = 0,
\end{eqnarray}
the curvature tensor $H_{\mu\nu\eta}$ remains invariant. Now, it is straightforward 
to check that the modified  FT-Lagrangian density (25) respects {\it both} the (1-form) 
YM gauge transformations (3) as well as the (2-form) tensorial gauge transformations (27) because
\begin{eqnarray}
\delta_g\, {\cal L}_{(M)} = 0,  \quad \delta_t\, {\cal L}_{(M)} 
= -\,\partial_\mu \left[\frac{m}{2} \varepsilon^{\mu\nu\eta\kappa} 
(A_\nu + \phi_\nu) \cdot [(A_\eta + \Phi_{\eta}) \times \Lambda_\kappa ] \right],
\end{eqnarray}
which imply that the action integral $S = \int d^4x \, {\cal L}_{(M)}$ remains 
invariant for the physically well-defined fields that vanish off at infinity. 
It is to be noted that, in the above derivation, we have taken
$\delta_g\, K_\mu = - (K_\mu \times \Omega)$ in addition to (3) in the (1-form)
YM gauge symmetry transformations.
We would like  to emphasize that the antisymmetry properties  of $\varepsilon_{\mu\nu\eta\kappa}$
are heavily used in the computation of $\delta_t\, {\cal L}_{(M)}$. For the sake of reader's
convenience, we mention a few intermediate steps in our Appendix A in the explicit proof of (28)
which is quite straightforward.

We note that the curvature tensor (26) is similar in appearance as the following  curvature tensor [17] of 
the Lahiri-model (LM) of the dynamical 4D non-Abelian 2-form theory, namely; 
\begin{eqnarray}
H^{(L)}_{\mu\nu\eta} &=& D_\mu B_{\nu\eta} + D_\nu B_{\eta\mu} + D_\eta B_{\mu\nu}
 - (K_\mu \times F_{\nu\eta}) \nonumber\\
&-& (K_\nu \times  F_{\eta\mu})  -  (K_\eta \times  F_{\mu\nu}),
\end{eqnarray}
where the superscript $(L)$ denotes the curvature tensor for the LM and 
$D_\mu B_{\nu\eta} = \partial_\mu B_{\nu\eta} - (A_\mu \times B_{\nu\eta})$.
Thus, there is a difference in the definition of the covariant derivative ${\tilde D}_\mu$ (which
includes both the 1-form fields $A_\mu$ and $\phi_\mu$) and the covariant derivative $D_\mu$ that
incorporates {\it only} the non-Abelian 1-form field $A_\mu$. The other key difference is the 
topological mass term for the FT-model [cf. (24)] which is totally different in 
its appearance {\it vis-\`{a}-vis} the topological term of the LM. In fact 
the latter is equal to [$(m/4) \varepsilon^{\mu\nu\eta\kappa}\, B_{\mu\nu} \cdot F_{\eta\kappa}$]. 
Furthermore, the transformation properties of the topological terms, under the tensorial 
gauge symmetry transformations, 
are totally different in both these models (see, e.g. [17,19] for details).
The common feature is the observation that the Lagrangian density of {\it both} the models 
respect the 1-form YM as well as the 2-form
tensorial gauge symmetries (see, e.g. [17,19]). We would also like to point out that, in the proof of (28),
we do not need any help from the Bianchi identity (which is the case for 
the tensorial symmetry invariance of the original FT-model (cf. Sec. 2)).

\subsection{(Anti-)BRST Symmetries Corresponding to NYM Transformations: Superfield Approach}

From the structure of the Lagrangian density (25) and the definition of $H_{\mu\nu\eta}$ 
[cf. (26)], it is clear that the
modified version of FT-model is very similar to Lahiri-model [17] for which we have 
already performed thorough analysis within the framework
of augmented superfield formalism [19]. In what follows, we discuss some of the 
relevant issues that are needed for our paper
to be  self-contained. The method of calculation for FT-model would be same as in [19]. In this section, 
we shall {\it only} pin-point some of the relevant steps in the accurate derivation 
of the proper (anti-)BRST symmetries corresponding to the
tensorial gauge symmetry transformations.

We note that $\delta_t A_\mu = \delta_t f_{\mu\nu} = \delta_t \phi_\mu 
= \delta_t {\cal F}_{\mu\nu} = 0$  [cf. (4)]. As a consequence, these objects 
are {\it invariant} quantities under 
the tensorial gauge symmetry transformations.
Exploiting the key arguments of the augmented version of BT-superfield approach, 
it is evident that the counterparts  of $A_\mu(x),\, \phi_\mu(x), \, f_{\mu\nu},\, F_{\mu\nu},\,
 {\cal F}_{\mu\nu}$ on the (4, 2)-dimensional supermanifold {\it must}
be independent of the ``soul'' coordinate $\theta$ and $\bar\theta$. 
Thus, we have the following relations:
\begin{eqnarray}
{\tilde A}^{(1)} =  A^{(1)}, \quad {\tilde \Phi}^{(1)} = \Phi^{(1)}, \quad 
{\tilde f}^{(2)} = f^{(2)}, \quad \tilde{ F}^{(2)} = { F}^{(2)},
\quad \tilde{\cal F}^{(2)} = {\cal F}^{(2)},
\end{eqnarray}
where all the notations have been explained in our Sec. 3.
As a consequence of the above equality,  we obtain the following relationships:
\begin{eqnarray}
&& \partial_\mu \Big({\tilde {\cal B}}^{(g)}_\nu + {\tilde \Phi}^{(g)}_\nu \Big) 
- \partial_\nu \Big ( {\tilde {\cal B}}^{(g)}_\mu 
+ {\tilde \Phi}^{(g)}_\nu \Big) \nonumber\\
&&+ i \,\Bigl[\Big({\tilde {\cal B}}^{(g)}_\mu 
+ {\tilde \Phi}^{(g)}_\mu \Big), \, \Big( {\tilde {\cal B}}^{(g)}_\mu  
+ {\tilde \Phi}^{(g)}_\mu \Big) \Bigr] = {\cal F}_{\mu\nu}(x), \nonumber\\
&& \tilde {\cal B}^{(g)}_\mu (x, \theta, \bar\theta) = A_\mu (x),  \quad
\partial_\mu {\tilde \Phi}^{(g)}_\nu - \partial_\nu {\tilde \Phi}^{(g)}_\mu 
+ i\, \Bigl[ {\tilde \Phi}^{(g)}_\mu, \, {\tilde \Phi}^{(g)}_\mu \Bigr] 
= f_{\mu\nu}(x),  \nonumber\\ 
&& {\tilde \Phi}^{(g)}_\mu (x, \theta, \bar\theta) = \phi_\mu (x), 
\quad \partial_\mu {\tilde {\cal B}}^{(g)}_\nu 
- \partial_\nu {\tilde {\cal B}}^{(g)}_\mu + i\, \Bigl[{\tilde {\cal B}}^{(g)}_\mu, 
\, {\tilde {\cal B}}^{(g)}_\mu \Bigr] = F_{\mu\nu}(x),
\end{eqnarray}
where the superscript ($g$) stands for the superfields obtained after the GIRs (30).
It is clear from the above equation that the gauge fields $A_\mu$ and $\phi_\mu$ 
and their corresponding curvature tensors
$F_{\mu\nu},\, f_{\mu\nu}$ and ${\cal F}_{\mu\nu}$ do {\it not} change {\it at all} 
under the proper (anti-)BRST transformations
corresponding to the tensorial gauge symmetry transformations.

Now the crucial observation is the fact that $\delta_t\, H_{\mu\nu\eta} = 0$ which shows 
that the curvature tensor is a gauge {\it invariant} quantity. As a consequence, 
within the framework of augmented superfield formalism, we shall
obtain the equality of $\tilde{\cal H}^{(3)} = H^{(3)}$ where the r.h.s. and l.h.s. 
of this HC are explicitly expressed as:
\begin{eqnarray}
H^{(3)} &=& d\,B^{(2)} + i\;\Big(\{A^{(1)} + \phi^{(1)}\} \wedge B^{(2)} 
- B^{(2)} \wedge \{A^{(1)} + \phi^{(1)}\} \Big) \nonumber\\ 
&+& i\;\Big( K^{(1)} \wedge {\cal F}^{(2)} - {\cal F}^{(2)} \wedge  K^{(1)}\Big), \nonumber\\
\tilde {\cal H}^{(3)} &=& \tilde {d} \;\tilde {\cal B}^{(2)}+ i\; \Big( \{A^{(1)} 
+ \phi^{(1)}\} \wedge \tilde {\cal B}^{(2)} 
- \tilde {\cal B}^{(2)} \wedge \{A^{(1)} + \phi^{(1)}\} \Big) \nonumber\\
&+& i \;\Big(\tilde {\cal K}^{(1)} \wedge {\cal F}^{(2)} - {\cal F}^{(2)} 
\wedge \tilde {\cal K}^{(1)}\Big).
\end{eqnarray}
We point out that we have taken  the ordinary $A^{(1)}$ and 
$\phi^{(1)}$ in $\tilde {\cal H}^{(3)}$ because of the fact that restrictions 
(30) are true. The ordinary curvature 2-form ${\cal F}^{(2)} = d\, \left(A^{(1)} +  \Phi^{(1)}\right) + i \;\left(A^{(1)} 
+ \Phi^{(1)}\right) \wedge \left(A^{(1)} + {\Phi}^{(1)}\right)$  and its supersymmetric version 
$ \tilde {\cal  F}^{(2)}$ (cf. (6)) have 
already been defined in Sec. 3 and $\tilde {\cal K}^{(1)}$ and ${\tilde{\cal B}}^{(2)}$ 
are as follows
\begin{eqnarray}
\tilde {\cal K}^{(1)}&=& dx^{\mu}\;\tilde {\cal K}_{\mu}(x, \theta, \bar\theta) 
+ d\theta \;{\tilde {\bar{\cal F}}}_{1}(x, \theta, \bar\theta) 
+ {d\bar\theta} \;\tilde {\cal F}_{1}(x,\theta,\bar\theta), \nonumber\\
\tilde {\cal B}^{(2)}&=&\frac{1}{2!}\; (dZ^M \wedge dZ^N)\;\tilde {\cal B}_{MN}(x, 
\theta,\bar\theta)\nonumber\\
&\equiv & \frac{1}{2!} \;(dx^{\mu}\wedge dx^{\nu})\; 
\tilde {\cal B}_{\mu \nu}(x, \theta,\bar\theta) 
+(dx^{\mu} \wedge d\theta)\; {\tilde {\bar{\cal F}}}_{\mu}(x,\theta,\bar\theta)\nonumber\\
&+&(dx^\mu \wedge d\bar\theta) \;\tilde {\cal F}_\mu (x,\theta,\bar\theta)
+(d\theta \wedge d\bar\theta)\;\tilde\Phi(x,\theta,\bar\theta)\nonumber\\
&+&(d\theta \wedge d\theta)\;{\tilde{\bar{\cal \beta}}}(x,\theta,\bar\theta) 
+ (d\bar\theta \wedge d\bar\theta)\;\tilde {\cal \beta}(x,\theta,\bar\theta),
\end{eqnarray}
where the expansions have been taken along the Grassmannian directions of 
the (4, 2)-dimensional supermanifold. The crucial difference between $H^{(3)}_{(L)}$ 
of LM (see, e.g. [17]) and $H^{(3)}$ of our discussion 
is the replacement of $A_\mu$ by ($A_\mu + \phi_\mu$) and $F_{\mu\nu}$ by ${\cal F}_{\mu\nu}$.
This is why, one observes that the equation (32) is the analogue of $H^{(3)}_{(L)}$ of LM [17]
but with the replacements: $A^{(1)}\,\rightarrow \; (A^{(1)} + \phi^{(1)})$ and
${\tilde F}^{(2)} \;\rightarrow \; \tilde{\cal F}^{(2)}$.

The supermultiplet superfields in (33) can be expanded along the Grassmannian directions 
$\theta$ and $\bar\theta$ of (4, 2)-dimensional supermanifold  as follows:
\begin{eqnarray}
\tilde {\cal B}_{\mu\nu} (x, \theta, \bar \theta) &=& B_{\mu\nu} (x) 
+ \theta \;\bar R_{\mu\nu} (x) + \bar \theta \;R_{\mu\nu} (x)
+ i \;\theta \;\bar\theta \;S_{\mu\nu} (x), \nonumber\\
\tilde {\cal F}_{\mu}(x, \theta, \bar\theta) &=& C_\mu (x) 
+ \theta \;{\bar B}^{(1)}_\mu(x) 
+ \bar \theta \; B^{(1)}_\mu(x) +i\; \theta\;\bar\theta \;S_\mu(x), \nonumber\\
{\tilde {\bar{\cal F}}}_{\mu}(x, \theta, \bar\theta) &=& {\bar C}_\mu (x)+ \theta \; \bar B^{(2)}_\mu(x) 
+ \bar \theta \; B^{(2)}_\mu(x) +i\; \theta\;\bar\theta \; {\bar S}_\mu(x),  \nonumber\\
\tilde {\cal K}_\mu (x, \theta, \bar \theta) &=& K_\mu (x) + \theta \; \bar P_\mu (x) + \bar\theta \; 
P_\mu (x) + i\;\theta \;\bar\theta\;Q_\mu (x),\nonumber\\
\tilde \Phi(x, \theta, \bar\theta)&=& \phi(x) + \theta \; {\bar f}_1(x) + \bar\theta \; f_1(x) 
+ i\; \theta\;\bar\theta \; b_1(x), \nonumber \\
\tilde \beta (x, \theta, \bar\theta)&=& \beta(x) + \theta \;{\bar f}_2(x) + \bar\theta \; f_2(x) 
+ i\; \theta\;\bar\theta \; b_2(x), \nonumber\\
\tilde {\bar \beta} (x, \theta, \bar\theta)&=& {\bar \beta}(x) + \theta \; 
{\bar f}_3(x) + \bar\theta \; f_3(x) 
+ i\; \theta\;\bar\theta \; b_3(x), \nonumber\\
\tilde {\cal F}_1 (x, \theta, \bar\theta)&=& C_1(x) + i\;\theta \; {\bar R}(x) 
+ i\;\bar\theta \; R(x) 
+ i\; \theta\;\bar\theta \; s_1 (x), \nonumber\\
{\tilde {\bar{\cal F}}}_1 (x, \theta, \bar\theta)&=& {\bar C_1}(x) + i\;\theta \; {\bar S}(x) 
+ i\;\bar\theta \; S(x) 
+ i\; \theta\;\bar\theta \;{\bar s}_1 (x),
\end{eqnarray}
where all the secondary fields on the r.h.s. would be expressed in terms of the 
basic and auxiliary  fields of the ordinary 4D BRST invariant Lagrangian density
due to HC (${\tilde {\cal H}}^{(3)} = H^{(3)}$).
In the above, we have (${\bar R}_{\mu\nu},\, R_{\mu\nu},\, {\bar P}_\mu,\,
P_\mu,\, {\bar S}_\mu, S_\mu,\, {\bar f}_1,$ $f_1, {\bar f}_2,\, f_2, 
{\bar f}_3, \, f_3, \, {\bar s}_1,\, s_1$) and ($S_{\mu\nu}, Q_\mu,\,{\bar B}^{(1)}_\mu,
B^{(2)}_\mu, b_1,\,b_2,\, b_3,\, R, \,\bar R,\, S,\, \bar S$) secondary fields 
as the set of fermionic and bosonic fields, respectively.  We further note that, 
in the limit [$(\theta, \bar\theta) = 0$], we obtain the ordinary 
4D basic and auxiliary fields of our (anti-)BRST 
invariant theory which are nothing but $B_{\mu\nu},\, C_\mu, \bar C_\mu,\, 
K_\mu,\, \phi, \beta, \bar\beta,\, C_1,\, \bar C_1$
where the set ($B_{\mu\nu},\, K_\mu,\, \phi, \beta, \bar\beta$) are the bosonic fields and  the set 
($C_\mu, \,\bar C_\mu,\, C_1, \,\bar C_1$) are the fermionic fields.

Taking the explicit forms of ${\tilde{\cal H}}^{(3)}$ and $H^{(3)}$ from (32), 
we can proceed along the same  lines as our earlier work on the augmented 
superfield approach to LM (see, e.g. [19] for details) 
and determine all the above fermionic and bosonic secondary fields 
in terms of ordinary basic and auxiliary fields.
We skip here the details of these calculations and point out the final 
expression for the (anti-)BRST
transformations as (see, e.g. [19]):
\begin{eqnarray}
&&s_{ab}B_{\mu\nu} = - \; ({\tilde D}_{\mu}\bar C_{\nu} -{\tilde D}_{\nu} \bar C_{\mu} ) 
+ \; \bar C_{1} \times {\cal F}_{\mu\nu},\; 
\qquad s_{ab}\bar C_{\mu} = - {\tilde D}_{\mu} \bar \beta, \nonumber\\
&& s_{ab}C_{\mu} = \bar B_{\mu},\; \qquad s_{ab} B_{\mu} = {\tilde D}_{\mu} {\rho},\;
\qquad s_{ab} C_1 = i\; \bar B_{1},\; \qquad s_{ab} \phi = -\rho, \nonumber\\
&& s_{ab} \bar C_1 = - \bar \beta,  \quad s_{ab} B_{1} = - i \; \rho, \;
\quad s_{ab} K_{\mu} = {\tilde D}_{\mu} \bar C_{1} - \bar C_{\mu}, \quad s_{ab} \beta 
= - \lambda,\nonumber\\ && s_{ab} [A_{\mu},\; \phi_\mu,\; F_{\mu\nu},
\; f_{\mu\nu},\; {\cal F}_{\mu\nu},\; H_{\mu\nu\eta},
 \bar \beta,  \; \bar B_1, \; \rho,\;  \lambda, \; \bar B_{\mu} ] = 0,
\end{eqnarray}
\begin{eqnarray}
&&s_{b} B_{\mu\nu} = - \; ({\tilde D}_{\mu}C_{\nu} - {\tilde D}_{\nu}C_{\mu} ) 
+ \; C_{1} \times {\cal F}_{\mu\nu},\;\qquad
s_{b} C_{\mu} =- {\tilde D}_{\mu}\beta, \nonumber\\ 
&&s_{b} \bar C_{\mu} = B_{\mu},\qquad s_b  {\bar B_1} = i\; \lambda, \qquad s_b \bar C_1 
= i\; B_{1},\qquad  s_{b} \bar B_{\mu} = - {\tilde D}_{\mu} \lambda, \nonumber\\ 
&& s_{b} K_{\mu} = {\tilde D}_{\mu}C_{1} - C_{\mu},\,\qquad s_{b} \phi = \lambda,\, 
\qquad s_b C_1 = -\beta,
\,\qquad s_{b} \bar \beta = \rho,\; \nonumber\\ && s_{b} [ A_{\mu}, \;\phi_\mu,\; 
F_{\mu\nu},\; f_{\mu\nu},\; {\cal F}_{\mu\nu},\;
H_{\mu\nu\eta},\; \beta, \; B_1, \; \rho,  \;\lambda, \; B_{\mu} ] = 0.
\end{eqnarray}
A close and careful look at the above transformations shows that these 
transformations are exactly same as the {\it ones} derived in the case 
of LM except we have the following replacements (see, e.g. [19] for details) 
\begin{eqnarray}
D_\mu \; \longrightarrow \; {\tilde D}_\mu,\qquad\qquad F_{\mu\nu} 
\;\longrightarrow \; {\cal F}_{\mu\nu}.
\end{eqnarray}
In other words, we observe that the (anti-)BRST symmetry transformations, corresponding to the (2-form) 
tensorial (NYM) gauge symmetry transformations, for the modified FT-model and LM of dynamical 2-form
non-Abelian gauge theory are connected with each-other through the above replacements (cf. (37)).
It is clear that the LM is a limiting case ($\phi_\mu = 0$) of the modified version of FT-model at the
{\it quantum level} where the nilpotent and absolutely anticommuting (anti-)BRST symmetries exist.

We further  note that the (anti-)BRST symmetry transformations are off-shell nilpotent
 ($s^2_{(a)b} = 0$) of order two and they are absolutely anticommuting 
($s_b\, s_{ab} + s_{ab}\, s_b = 0$) on the constraint hyper-surface
defined by the following field equations:
\begin{eqnarray}
B_\mu + \bar B_\mu + {\tilde D}_\mu \phi = 0,\quad B + \bar B + i (C \times \bar C) = 0,
 \quad B_1 + {\bar B}_1 - i \phi = 0.
\end{eqnarray}
The above hyper-surface is embedded in the 4D flat Minkowskian spacetime manifold and it 
is described by the above Curci-Ferrari (CF) type restrictions (38). 
We note that these field equations 
are found to be (anti-)BRST invariant. The CF-condition ($B + \bar B 
+ i(C \times \bar C) = 0$) remains invariant under the (anti-)BRST 
symmetry corresponding to the (1-form) YM gauge symmetries and the other two
CF-type conditions are invariant under the nilpotent (anti-)BRST symmetry transformations 
corresponding to the NYM gauge symmetries
(cf. (35, 36)). As pointed out earlier, the CF-type conditions are the signatures of a gauge theory when 
it is discussed within the framework of BRST formalism which primarily imply the 
independent nature of anti-BRST symmetries.

By exploiting the standard tricks of the (anti-)BRST symmetries, one can obtain the (anti-) BRST invariant 
Lagrangian densities (see, e.g. [19] for details):
\begin{eqnarray}
{\cal L}_{\bar B_1}  &=& \frac{1}{12} \,H^{\mu\nu\eta}\cdot H_{\mu\nu\eta}
- \frac{1}{4} \,F^{\mu\nu}\cdot F_{\mu\nu} 
+ \frac{m^2}{2}\; \phi_\mu \cdot \phi^\mu \nonumber\\ 
&-&\frac{m}{2}\, \varepsilon^{\mu\nu\eta\kappa}\, (A_\nu + \phi_\nu) 
\cdot \Big[ {\tilde D}_\mu B_{\eta\kappa}
+ \frac{1}{2}\, (A_\mu + \phi_\mu) \times B_{\eta\kappa} \Big] 
+ {\bar B}^{\mu} \cdot {\bar B}_\mu  \nonumber\\
&+& \frac{i}{2}\; B^{\mu\nu} \cdot ({\bar B}_1 \times {\cal F}_{\mu\nu}) 
+ ({\tilde D}_\mu B^{\mu\nu} + {\tilde D}^{\nu}\phi) \cdot {\bar B}_\nu 
+ {\tilde D}_\mu \bar \beta \cdot {\tilde D}^{\mu} \beta \nonumber\\ 
&+& \frac{1}{2} \;\Bigl [({\tilde D}_\mu \bar C_\nu - {\tilde D}_\nu \bar C_\mu ) 
- \bar C_1 \times {\cal F}_{\mu\nu} \Bigr ] \cdot \Bigl [({\tilde D}^\mu C^\nu 
- {\tilde D}^\nu  C^\mu ) 
- C_1 \times {\cal F}^{\mu\nu} \Bigr ]\nonumber\\
&+& \rho \cdot ({\tilde D}_\mu C^\mu - \lambda ) +  ({\tilde D}_\mu {\bar C}^\mu 
- \rho )\cdot \lambda,\nonumber\\
{\cal L}_{B_1}  &=& \frac{1}{12}\, H^{\mu\nu\eta}\cdot H_{\mu\nu\eta}
- \frac{1}{4} \,F^{\mu\nu}\cdot F_{\mu\nu} 
+ \frac{m^2}{2} \,\phi_\mu \cdot \phi^\mu \nonumber\\ 
&-&\frac{m}{2}\, \varepsilon^{\mu\nu\eta\kappa} (A_\nu + \phi_\nu) 
\cdot \Big[ {\tilde D}_\mu B_{\eta\kappa}
+ \frac{1}{2} \,(A_\mu + \phi_\mu) \times B_{\eta\kappa} \Big] 
+ B^{\mu} \cdot B_\mu  \nonumber\\
&-& \frac{i}{2}\; B^{\mu\nu} \cdot (B_1 \times {\cal F}_{\mu\nu})
- ({\tilde D}_\mu B^{\mu\nu} - {\tilde D}^{\nu}\phi) \cdot B_\nu 
+ {\tilde D}_\mu \bar \beta \cdot {\tilde D}^{\mu} \beta \nonumber\\
 &+& \frac{1}{2} \;\Bigl [({\tilde D}_\mu \bar C_\nu - {\tilde D}_\nu \bar C_\mu ) 
- \bar C_1 \times {\cal F}_{\mu\nu} \Bigr ] \cdot \Bigl [({\tilde D}^\mu C^\nu 
- {\tilde D}^\nu C^\mu ) - C_1 \times {\cal F}^{\mu\nu} \Bigr ]\nonumber\\
&+& \rho \cdot ({\tilde D}_\mu C^{\mu} - \lambda ) +  ({\tilde D}_\mu {\bar C}^{\mu} 
- \rho )\cdot \lambda, 
\end{eqnarray}
which remain invariant under the (2-form) tensorial nilpotent  (anti-)BRST symmetry 
transformations $s_{(a)b}$ listed in (35, 36) because we have
\begin{eqnarray}
&&s_{ab}\, {\cal L}_{{\bar B}_1}  =  -\,\partial_\mu\, \Bigl[\frac{m}{2}\, \varepsilon^{\mu\nu\eta\kappa} 
(A_\nu + \phi_\nu) \cdot \{(A_\eta + \Phi_{\eta}) \times \bar C_\kappa 
+ {\bar C}_1 \times {\cal F}_{\eta\kappa}\} \nonumber\\ &&+ \rho \cdot {\bar B}^\mu 
+  \lambda \cdot {\tilde D}^\mu {\bar \beta}
+ ({\tilde D}^\mu {\bar C}^\nu - {\tilde D}^\nu {\bar C}^\mu ) \cdot {\bar B}_\nu 
- ({\bar C}_1 \times {\cal F}^{\mu\nu} ) \cdot {\bar B}_\nu \Bigr],\nonumber\\
&& s_b\,{\cal L}_{B_1}  =  \,\partial_\mu\, \Bigl[-\,\frac{m}{2}\, \varepsilon^{\mu\nu\eta\kappa} 
(A_\nu + \phi_\nu) \cdot \{(A_\eta + \Phi_{\eta}) \times C_\kappa + { C}_1 \times {\cal F}_{\eta\kappa}\}
 \nonumber\\  &&+ \rho\cdot {\tilde D}^\mu \beta + \lambda \cdot B^\mu 
+ ({\tilde D}^\mu C^\nu - {\tilde D}^\nu C^\mu ) \cdot B_\nu 
-  (C_1 \times {\cal F}^{\mu\nu} ) \cdot  B_\nu \Bigr].
\end{eqnarray}
We close this section with the remark that (anti-)BRST charges, ghost conserved  
charge, etc., can be computed by exploiting the standard tricks
of Noether's theorem exactly like what we have done [19] for the LM. 
Similarly, the standard BRST algebra can be computed along the same line as performed in [19].

\section{Conclusions}

The  central results of our present investigation are the derivations of proper  
(anti-) BRST symmetries corresponding to the (N)YM gauge symmetries of
the FT-model. We have derived the  (anti-)BRST symmetries, corresponding 
to the (1-form) YM gauge symmetries by exploiting the theoretical tricks that 
are distinctly different from the {\it ones}
adopted in our previous endeavor [18]. The derivation of the proper (anti-)BRST 
symmetries for the NYM gauge symmetries was a {\it challenging} problem for  
us (within the framework of the augmented version of  BT-superfield formalism).
In the accomplishment of the latter goal, we have been theoretically compelled to 
incorporate an auxiliary field and the kinetic term for the 2-form gauge field. 
As a consequence, the 2-form field becomes {\it dynamical} (even though this field happens 
to be an auxiliary field in the {\it original} FT-model). This is a {\it new} observation 
in our present investigation (where the Lagrangian density is modified due to theoretical needs.

We are theoretically forced to go beyond the HC to derive the full set 
of proper off-shell nilpotent (anti-)BRST symmetry transformations
by invoking the appropriate GIRs in the context of  (1-form) 
YM gauge symmetry transformations. This is {\it also}
a novel observation. However, in the case of the derivation of proper 
(anti-)BRST symmetry transformations (corresponding to the (2-form) 
tensorial (NYM) gauge symmetry transformations), we invoke {\it only} the HC
(i.e. $\tilde {\cal H}^{(3)} = H^{(3)}$) for the modified version of 
the FT-model where the curvature tensor $H_{\mu\nu\eta}$ is defined in (26). 
The  computations are similar  in texture and contents as is the case with  
the dynamical non-Abelian 2-form gauge theory of LM which  we have derived within the framework of superfield formalism [19]. 
Thus, we have quoted these results with suitable modifications in our present endeavor.

It is worthwhile to point out that the modified version of FT-model, at the classical level, is essential only because
of the fact that we wish to derive the proper (anti-)BRST symmetry transformations corresponding to the tensorial gauge
symmetry transformations (4) within the framework of the augmented version of BT-superfield formalism where HC 
(i.e. $\tilde {\cal H}^{(3)} = H^{(3)}$) plays a very crucial role (cf. (32)). The original FT-model has been studied
in an excellent set of papers [23-27] and its different aspects have been discussed within the framework of BRST, 
Batalin-Vilkovisky (antifield-antibracket), 't Hooft-Veltman, etc., formalisms. In these works [23-27], the Batalin-Vilkovisky
formalism has been applied to obtain the proper (anti-)BRST symmetries, one-loop unitarity of this theory has been proven
by using the Ward-identities and 't Hooft-Veltman procedure, quantization of the charged 2-form and string fields have been
performed, similarities and differences with the Witten's string field theory have been pointed out. None of these
formalisms, however, have applied the strength of HC to derive the nilpotent and absolutely anticommuting (anti-)BRST symmetry
transformations which is the central theme of our present investigation within the framework of BT-superfield formalism. 
This is why, in these works [23-27], there has been no need of the modification of the FT-model which have been performed in
our present work. The outcome of our present endeavor should be seen in the emergence of the new ``mass'' term in
the Lagrangian density (cf. (25)) which is found to be invariant under the YM and NYM gauge symmetries.

The differences between FT-model and LM have been pointed out in our present 
investigation. The modified version of FT-model becomes dynamical 
non-Abelian 2-form gauge theory similar to the dynamical theory considered
in the case of LM. However, the structure of the topological mass term is totally different
in both these cases. The common feature is the observation that, in
{\it both} these theories, the topological mass term remains invariant under both the appropriate 
(N)YM symmetry transformations of the respective theories. 
We have also demonstrated that, at the quantum level (where the (anti-)BRST 
symmetries are valid), the LM is a liming case of the modified version of 
FT-model when we apply the augmented version of BT-superfield approach to
BRST formalism. It would be a nice idea to apply our
augmented version of BT-superfield formalism to other non-Abelian 
higher $p$-form ($p \ge 3$) gauge theories and study their novel features. 
We plan to pursue this direction of investigation in our future
endeavors [28].\\

\noindent
{\bf Acknowledgements} \\

\noindent
Two of us (AS and SK) would like to gratefully acknowledge the financial support
from CSIR  and UGC , Government of India, New Delhi, under their SRF-schemes.\\


\section*{\hspace {4.9 cm} \large{\bf{Appendix A}}}

\vspace{0.5 cm}
\noindent
Here we provide some of the intermediate steps to show that the modified 
FT-Lagrangian density (25) transforms to a total spacetime derivative under 
the tensorial (2-form) gauge symmetry transformations (27). First of all, it can be
checked that $\delta_t\, H_{\mu\nu\eta} = 0$ because of the following basic definition:
\begin{eqnarray}
[{\tilde D}_\mu,\, {\tilde D}_\nu]\, \Lambda_\eta = - ({\cal F}_{\mu\nu} \times \Lambda_\eta),
\qquad {\tilde D}_\mu\, \Lambda_\eta = \partial_\mu \,\Lambda_\eta - (A_\mu + \phi_\mu)\times \Lambda_\eta,
\end{eqnarray}
where $[{\tilde D}_\mu,\, {\tilde D}_\nu] = {\tilde D}_\mu\, {\tilde D}_\nu - {\tilde D}_\nu 
\, {\tilde D}_\mu$ is the
commutator. Next, we note that topological term transforms under $\delta_t$ as:
\begin{eqnarray}
\delta_t \Big[ - \frac{m}{2}\, \varepsilon^{\mu\nu\eta\kappa}(A_\nu 
+ \Phi_\nu) \cdot {\tilde D}_\mu B_{\eta\kappa} 
- \frac {m}{4}\varepsilon^{\mu\nu\eta\kappa}\,(A_\nu + \Phi_\nu) \cdot [(A_\mu 
+ \Phi_\mu) \times B_{\eta\kappa}] \Big],
\end{eqnarray}
where there are two terms in the square bracket. Using equation (26) and the 
antisymmetric property of $\varepsilon_{\mu\nu\eta\kappa}$,
we observe that the first term transforms as: 
\begin{eqnarray}
\delta_t \Big[ - \frac{m}{2}\, \varepsilon^{\mu\nu\eta\kappa}(A_\nu 
+ \Phi_\nu) \cdot {\tilde D}_\mu B_{\eta\kappa} \Big]
= - \frac{m}{2}\,\varepsilon^{\mu\nu\eta\kappa} (A_\nu 
+ \Phi_\nu) \cdot ({\cal F}_{\mu\eta} \times \Lambda_\kappa).
\end{eqnarray}
In the computation of the transformation on the second term, 
we note that $\delta_t\, B_{\eta\kappa} = 
{\tilde D}_\eta \Lambda_\kappa - {\tilde D}_\kappa \Lambda_\eta$ 
which leads to a single term due to presence of the Levi-Civita tensor.
That is to say, we have the following explicit transformation:
\begin{eqnarray}
\delta_t [- \frac{m}{4}\varepsilon^{\mu\nu\eta\kappa}\,(A_\nu + \Phi_\nu) \cdot [(A_\mu 
+ \Phi_\mu) \times B_{\eta\kappa}] \nonumber\\
= + \frac{m}{2} \varepsilon^{\mu\nu\eta\kappa}\, (A_\nu + \Phi_\nu) \cdot [(A_\mu 
+ \Phi_\mu) \times {\tilde D}_\eta \Lambda_\kappa].
\end{eqnarray}
The above expression can be written as the sum of two terms due to the fact that:
 ${\tilde D}_\eta \Lambda_\kappa = \partial_\eta \Lambda_\kappa 
- (A_\eta + \Phi_\eta) \times \Lambda_\kappa$. Finally, we find that it can be expressed as follows:
\begin{eqnarray}
\partial_\eta \Big[ \frac{m}{2}\, \varepsilon^{\mu\nu\eta\kappa}\, (A_\nu 
+ \Phi_\nu) \cdot [(A_\mu + \Phi_\mu) \times \Lambda_\kappa ]\Big] \nonumber\\
+ \frac{m}{2}\,\varepsilon^{\mu\nu\eta\kappa}\, (A_\nu + \Phi_\nu) \cdot 
({\cal F}_{\mu\eta} \times \Lambda_\kappa).
\end{eqnarray}
Adding (43) and (45), it is straightforward to see that, under the 
tensorial gauge symmetry transformations, the modified Lagrangian density 
for the FT-model transforms to a total spacetime derivative which has been mentioned
in the main body of our text [cf. (28)] thereby making the action integral 
$S = \int d^4 x \, {\cal L}_{M}$ invariant.\\

\section*{\hspace {4.9cm} \large{\bf{Appendix B}}}

\vspace{0.5 cm}
\noindent
In this Appendix, we provide arguments in favor of our expressions in (7)
where we have  associated ghost fields {\it only} with the 1-form ($A^{(1)} = dx^\mu\, A_\mu$)
potential $A_\mu$ but we have {\it not} associated any ghost fields with 1-form field $\phi_\mu$.
Towards this goal in mind, let us have the general expansions for the 1-form
$\phi^{(1)} =dx^\mu \, \phi_\mu$, too. In other words, we have the following generalizations on the 
(4, 2)-dimensional supermanifold:   
\begin{eqnarray}
A^{(1)} \rightarrow \tilde A^{(1)} = dx^\mu \,\tilde{\cal B}_\mu (x, \theta, \bar\theta)
+ d\theta \, {\bar{ F}}_1 (x, \theta, \bar\theta)
+  d\bar\theta \, {F}_1 (x, \theta, \bar\theta), \nonumber\\
\Phi^{(1)} \rightarrow  \tilde\Phi^{(1)} = dx^\mu \,\tilde\Phi_\mu (x, \theta, \bar\theta)
+ d\theta \, {\bar {F}}_2 (x, \theta, \bar\theta)
+  d\bar\theta \,{ F}_2 (x, \theta, \bar\theta).
\end{eqnarray}
Now, we have the expansions of the superfields along the Grassmannian directions ($\theta, \bar\theta$)
of the (4, 2)-dimensional supermanifolds  as listed below
\begin{eqnarray}
\tilde {\cal B}_\mu (x, \theta, \bar\theta) &=& A_\mu (x) + \theta \;{\bar R}_\mu (x) + \bar\theta \; R_\mu (x) 
 + i\; \theta \,\bar\theta\; P_\mu (x),\nonumber\\
\tilde \Phi_\mu (x, \theta, \bar\theta) &=& \phi_\mu (x) + \theta \; {\bar S}_\mu^{(1)} (x) 
+ \bar\theta \;S_\mu^{(1)} (x)  
+i \;\theta\, \bar\theta \;T_\mu (x), \nonumber\\ 
{{ F}}_1 (x, \theta, \bar\theta) &=& C_1 (x) + \theta \; {\bar B}_3 (x) + \bar\theta \; B_3 (x) 
 +i \;\theta \bar\theta \;s_3 (x), \nonumber\\
{ {\bar {F}}}_1(x, \theta, \bar\theta ) &=& {\bar C}_1 (x) 
+ \theta \; {\bar B}_4 (x) + \bar\theta \;B_4 (x)  
+i \;\theta\, \bar\theta \;{\bar s}_3(x), \nonumber\\
{ F}_2 (x, \theta, \bar\theta) &=& C_2 (x) + \theta \; \bar R_1 (x) + \bar\theta \;R_1 (x) 
 +i \;\theta \bar\theta \;s_4 (x), \nonumber\\
{\bar { F}}_2 (x, \theta, \bar\theta ) &=& {\bar C}_2 +(x) + \theta \; \bar S_1 (x) + \bar\theta \;S_1 (x)  
+i \;\theta\, \bar\theta \;{\bar s}_4 (x),
\end{eqnarray}
where the bosonic fields ($A_\mu, \phi_\mu, P_\mu, T_\mu, {\bar B}_3, B_3, {\bar B}_4, B_4, 
\bar R_1, R_1, \bar S_1, S_1$) and the fermionic fields ($C_1, {\bar C}_1, C_2, {\bar C}_2,
 R_\mu, {\bar R}_\mu,  S_\mu^{(1)}, {\bar S}_\mu^{(1)},  {s}_3, {\bar s}_3, {s}_4, {\bar s}_4$)
in the whole expansions match showing the validity of SUSY in the theory. It will be noted that the secondary 
fields on the r.h.s. of expansions in (47) are functions  of the 4D spacetime coordinate ($x^\mu$)
only.

We exploit the following (HC) (i.e. $\tilde {\cal F}^{(2)} = {\cal F}^{(2)}$), namely; 
\begin{eqnarray}
&& \tilde d\, \left(\tilde A^{(1)} + \tilde \Phi^{(1)} \right) + i\; \left(\tilde A^{(1)} 
+ \tilde \Phi^{(1)}\right) \wedge \left(\tilde A^{(1)} + {\tilde \Phi}^{(1)}\right) \nonumber\\
&& = d\, \left(A^{(1)} +  \phi^{(1)} \right) + i\; \left(A^{(1)} 
+  \phi^{(1)}\right) \wedge \left(A^{(1)} +  \phi^{(1)}\right).
\end{eqnarray}
It will we noted that the r.h.s. of the 2-form contains only the spacetime differentials ($dx^\mu\wedge dx^\nu$)
but the l.h.s. incorporates $(dx^\mu\wedge dx^\nu),\, (dx^\mu\wedge d\theta),\, (dx^\mu\wedge d\bar\theta),\,
(d \theta\wedge d\bar\theta),\, (d \theta\wedge d\theta),\, (d \bar\theta\wedge d\bar\theta)$. 
 Due  the above  equality (48), we have to set the coefficients of 
$(dx^\mu\wedge d\theta),\, (dx^\mu\wedge d\bar\theta),\,
(d \theta\wedge d\bar\theta),\, (d \theta\wedge d\theta)$ and $(d \bar\theta\wedge d\bar\theta)$ equal to zero.
These restrictions yield the following useful relationships:  
\begin{eqnarray}
&& \partial_\mu \left( {\bar { F }}_1 +  {\bar { F }}_2 \right)
- \partial_\theta \left( \tilde {\cal B}_\mu + \tilde \Phi_\mu \right)  + i \,\left[ \tilde {\cal B}_\mu 
+ \tilde \Phi_\mu, \, {\bar { F }}_1 + {\bar { F }}_2 \right] = 0, \nonumber\\
&& \partial_\mu\left({ F}_1 + { F}_2 \right)- \partial_{\bar\theta} \left( \tilde {\cal B}_\mu 
+ \tilde \Phi_\mu \right)  + i \, \left[ \tilde {\cal B}_\mu 
+ \tilde \Phi_\mu, \,{F}_1 + {F}_2 \right] = 0, \nonumber\\
&& \partial _\theta\left( {{ F}}_1 + {{ F}}_2 \right) 
+ \partial_{\bar\theta} \left({\bar {F}}_1 + { {\bar{ F}}}_2\right)
+ i \; \lbrace {{F}}_1 + {{F}}_2, { {\bar{ F}}}_1 
+ { {\bar{ F}}}_2 \rbrace =0, \nonumber\\
&& \partial_{\theta} \left({{\bar{ F}}}_1  + { {\bar{ F}}}_2 \right)
+ \frac {i}{2}\; \lbrace { {\bar{ F}}}_1  + {{\bar{ F}}}_2, 
{ {\bar{ F}}}_1  + { {\bar{ F}}}_2\rbrace  = 0,\nonumber\\
&& \partial_{\bar\theta} \left({ F}_1 + { F}_1 \right)
+ \frac {i}{2} \; \lbrace { F}_1 +  {F}_2 ,
 {F}_1 +  { F}_2 \rbrace = 0.
\end{eqnarray}
The above restrictions  would play very important roles in the determination of the fermionic 
(anti-)BRST symmetries as we briefly  mention below.

In these relationships [cf. (49)], if we substitute the super-expansions (47), 
we obtain the following (anti-)BRST symmetry 
transformations (that are nothing but the coefficient of the 
Grassmannian variables $(\theta, \bar\theta)$), namely; 
\begin{eqnarray}
&& s_{ab} A_\mu = D_\mu ({\bar C}_1 + {\bar C}_2), 
\quad s_{ab} ({\bar C}_1 + {\bar C}_2) = \frac {1}{2} \left[({\bar C}_1 
+ {\bar C}_2)\times ({\bar C}_1 + {\bar C}_2)\right], \nonumber\\
&& s_{ab} \phi_\mu = -\left[\phi_\mu \times ({\bar C}_1 + {\bar C}_2)\right],\qquad \qquad s_{ab} (C_1 + C_2) 
= i\, ({\bar B}_3 + \bar R_1), \nonumber\\ &&
s_{ab} (B_4 + S_1)  = - \left[(B_4 + S_1)  \times ({\bar C}_1 + {\bar C}_2)\right],\qquad 
s_{ab} ({\bar B}_3 + \bar R_1) = 0,\nonumber\\  
&&  s_{ab} B_{\mu\nu}= - \left[B_{\mu\nu} \times ({\bar C}_1 + {\bar C}_2)\right],\qquad
 s_b A_\mu = D_\mu (C_1 + C_2), \nonumber\\ && s_b \phi_\mu = - \left[\phi_\mu \times (C_1 + C_2)\right],
\quad  s_b (C_1 + C_2) = \frac {1}{2}\left[(C_1 + C_2) \times (C_1 + C_2)\right],\nonumber\\ &&
s_b ({\bar C}_1 + {\bar C}_2) = i (B_4+ S_1),\quad 
  s_b ({\bar B}_3 + \bar R_1)= - \left[({\bar B}_3 + \bar R_1) \times (C_1 + C_2)\right], \nonumber\\ &&
 s_b (B_4 + S_1) = 0, \qquad s_b B_{\mu\nu} = - \left[B_{\mu\nu} \times (C_1 + C_2)\right].
\end{eqnarray}
The above equation shows that if we take $ B_4 + S_1 = B, {\bar B}_3 + \bar R_1 = \bar B, C_1 + C_2 = C$ and 
${\bar C}_1 + {\bar C}_2 = \bar C$, we shall obtain all the (anti-)BRST symmetry transformations 
that have been obtained in our earlier work [18]. Furthermore, the above identifications {\it also}
yield the celebrated CF-condition $B+ \bar B + i \,(C \times \bar C) = 0$ for the 4D 
non-Abelian  1-form gauge theory. In particular, we wish to mention that it is the third relationship from the top
and/or bottom in (49) that yields the (anti-)BRST invariant CF-condition. 
In other words, the expressions (written in (7)) are good enough 
to yield the proper (i.e. off-shell nilpotent and absolutely anticommuting) (1-form) YM
(anti-)BRST symmetry transformations  for our present theory which have been quoted in 
Sec. 2 (and derived in Sec. 3) in explicit fashion.




\end{document}